# Nonlinear optical response in a ferromagnetic insulating manganite: $Pr_{0.8}Ca_{0.2}MnO_3$

A. Nakano [1*], K. Uchida[1], Y. Tomioka[2], M. Takaya[3], Y. Okimoto[4] and K. Tanaka[1,5*]

[1] Department of Physics, Graduate School of Science, Kyoto University, Kyoto, Kyoto 606-8502, Japan

[2] National Institute of Advanced Industrial Science and Technology (AIST), Tsukuba, 305-8565, Japan.

[3] Department of Geology and Mineralogy, Graduate School of Science, Kyoto University, Kyoto, Kyoto 606-8502, Japan

[4] Department of Chemistry, Tokyo Institute of Technology, Meguro, Tokyo 152-8551, Japan

[5] Institute for Integrated Cell-Material Sciences, Kyoto University, Sakyo-ku, Kyoto 606-8501, Japan

*e-mail: nakano.aiko.75n@kyoto-u.jp, koichiro.tanaka@rigen.jp

## Abstract

High harmonic generation from $Pr_{0.8}Ca_{0.2}MnO_3$ was investigated across a high-temperature paramagnetic phase and a low-temperature ferromagnetic phase. As the temperature decreases, the harmonic intensity gradually increases in the paramagnetic phase like that in different composition material $Pr_{0.6}Ca_{0.4}MnO_3$. However, it turns to a decrease in the ferromagnetic phase. We propose a possible interpretation of the anomaly around the ferromagnetic transition temperature considering the thermal fluctuation of orbital order and the metal-insulator phase separation in the ferromagnetic insulating phase.

## I. INTRODUCTION

High-harmonic generation (HHG) is an extremely nonlinear optical phenomenon that occurs when materials are irradiated by intense laser light [1,2]. It has been over a decade since the investigation of HHG extended to the various kinds of condensed materials from gas materials. In solids, HHG has emerged as a promising tool to extract material properties inaccessible through conventional linear optical measurements [3]. So far, HHG has mainly been investigated for materials in which one-electron approximation holds, and band theory can be applied.

There has been growing interest in the case that electron correlation is rather strong, and the local electron correlation effect has to be included to understand the HHG. Revealing the nature of the coupling of the local electronic correlation and HHG process will lead to a more sophisticated understanding of ultrafast carrier dynamics in strongly correlated materials under an intense laser field. Many theoretical studies have predicted unique aspects of HHG in strongly correlated materials [4–11], and a few experiments have been performed on them [12–15]. Our purpose is to clarify the key factors necessary for understanding the HHG properties of strongly correlated materials.

Hole-doped manganite $Pr_{1-x}Ca_xMnO_3$ is an example of strongly correlated materials with a coupling between spin, charge, and orbital (lattice) degrees of freedom, and it exhibits variations of electronic and magnetic states depending on temperature and hole doping level $x$ [16,17]. It is thus of importance to investigate changes of HHG respective to those of electronic/magnetic orderings in $Pr_{1-x}Ca_xMnO_3$, which may allow us to study a relation between HHG process and an electron correlation. In the hole-doped region of $0.3 \leq x < 0.8$, PCMO system undergoes $Mn^{3+}/Mn^{4+}$ charge ordering (CO) with orbital ordering (OO), and antiferromagnetic (AF) transition at lower temperature. In the very low-doped regime ($x \leq 0.25$), CO doesn't appear and only OO is observed [18]. The transition temperature $T_{OO}$ is lowered from ~950 to ~300 K as $x$ increases [16]. For $0.1 < x < 0.3$, a ferromagnetic (FM) insulating phase appears instead of AF. In the entire temperature range and the composition ratio including the FM ordered phase, PCMO system shows insulating resistivity [19]. The absence of metallicity or weaker double exchange (DE) interaction in PCMO is ascribed to the narrower bandwidth of the $e_g$-band, which comes from a larger tilting of $MnO_6$ octahedra.

Recently, we have reported a study on HHG from $Pr_{0.6}Ca_{0.4}MnO_3$ [15], which exhibits CO and AF transition below room temperature. As the temperature decreases, the HHG property changes at CO transition temperature, but no significant change occurs at AF transition temperature. As the temperature increases, the destructive interference among high-harmonic (HH) emissions from thermally activated multiple charge configurations is enhanced, which causes the reduction of HH intensity.

In the present study, $Pr_{0.8}Ca_{0.2}MnO_3$, whose composition is close to the CO region, was selected and examined under the same experimental conditions as $Pr_{0.6}Ca_{0.4}MnO_3$ to ensure the comparability of the results. In $Pr_{0.8}Ca_{0.2}MnO_3$, OO emerges at ~ 450 K, and with a further decrease in temperature, Mn magnetic moments align ferromagnetically at $T_{C1} \approx 130$ K, and Pr magnetic moments align ferromagnetically at $T_{C2} \approx 60$ K [16,20,21]. It should be noticed that there is spatial inhomogeneity below $T_{C1}$, where metallic-like regions are weakly linked to one another and become macroscopically insulating [22,23]. The presence or absence of reported spatial inhomogeneity in their ground states is an important difference between $Pr_{0.8}Ca_{0.2}MnO_3$ and $Pr_{0.6}Ca_{0.4}MnO_3$. Anomalous behavior was observed in the temperature dependence of HHG from $Pr_{0.8}Ca_{0.2}MnO_3$ near $T_{C1}$. This anomaly can be understood by considering thermal fluctuations of orbital order and spatial inhomogeneity in the ferromagnetic insulating phase.

## II. EXPERIMENT

HHG measurements were carried out in reflection geometry within the temperature range 6 K ≤ T ≤ 300 K, across paramagnetic (PM) phase to FM phase (indicated by red arrow in Fig. 1). Figure 2 (a) shows the schematic of the experimental setup. $Pr_{0.8}Ca_{0.2}MnO_3$ crystal was irradiated by intense mid-infrared (MIR) light pulse with a photon energy of 0.26 eV, repetition rate of 1 kHz, pulse duration of 100 fs, and intensity of 0.3 TW/cm$^2$ at the sample surface. The power of the incident MIR light was controlled by a pair of wire-grid polarizers (WG). By a combination of a quarter-wave plate (QWP) and WG, the polarization angle of the MIR light with respect to the sample crystal was controlled. The MIR light was focused on the sample by a reflective objective lens with the 20° angle of incidence. For more information on experimental setup, see Ref. [15].

Single crystal of $Pr_{0.8}Ca_{0.2}MnO_3$ was grown using floating zone techniques. The crystal rod was cut out and characterized by X-ray back-reflection Laue method. HHG measurements were performed with polarization of incident MIR light along one of the directions corresponding to the Mn-O bonds. The surface was prepared for reflection measurement by polishing and buffing, reducing surface roughness to less than 1 μm.

## III. RESULTS

Figure 2(b) presents typical HH spectra at 300 K, 125 K, and 25 K; above $T_{C1}$, around $T_{C1}$, and far below $T_{C1}$. We observed the clear third and fifth harmonic peaks at 0.8 eV and 1.3 eV, respectively. Even-order nonlinear optical processes are not allowed according to *Pbnm* symmetry of the material [16]. Both the third and fifth harmonics become maximum at $T_{C1} \approx 130$ K, i.e., the non-monotonic dependence on temperature, indicating a strong correlation with the ferromagnetic transition.

In Figs. 3(a) and 3(b), we show the detailed temperature variation of third and fifth harmonic intensities. With decreasing temperature, the harmonic intensities gradually increase above $T_{C1}$ and rapidly decrease just below $T_{C1}$. No significant change is observed at T = $T_{C2}$. As shown in Fig. 3(c), in the temperature range of 6 K ≤ T ≤ 300 K, optical gap energy is always relatively larger than the photon energy of incident light [19]. Therefore, the one-photon resonant enhancement does not occur at any temperature. The temperature variation of reflectivity at 0.26 eV shows a monotonic change, which increases by about 10% at low temperatures (see Supplemental Material [24]). Therefore, the experimental results cannot be explained only by the temperature dependence of linear optical response.

HH intensity dependence on MIR light intensity was measured at several temperatures around $T_{C1}$ (Figs. 4(a) and 4(b)). The deviation from the power low is observed both for third and fifth harmonics, clearly indicating that HHG is beyond the perturbation regime. In the HH intensity dependence on MIR light intensity, the weak monotonic temperature variation is observed, indicating that the mechanism of HHG remains almost unchanged around $T_{C1}$.

## IV. DISCUSSION

The temperature dependence of HH yields above $T_{C1}$ resembles the one we observed from other strongly correlated materials such as $Ca_2RuO_4$ [14] and $Pr_{0.6}Ca_{0.4}MnO_3$ [15]. In the previous studies, they have attributed the temperature-varying HH properties to the thermal fluctuation of the spin or charge order that strongly couples to the carrier dynamics. A similar phenomenology applies to $Pr_{0.8}Ca_{0.2}MnO_3$. It seems reasonable to assume that the HH intensity gradually increases as the temperature decreases, reflecting the strong coupling between the thermal fluctuation of orbital order and the carrier dynamics below $T_{OO} \sim 450$ K.

Let us now turn to discuss the sharp reduction of the harmonic intensities below $T_{C1}$. Following the above discussion, one might expect that the thermal fluctuation of the spin order is suppressed as the temperature decreases, and thus, the harmonic intensity is enhanced at low temperatures. However, the opposite temperature dependence was observed. We consider this behavior to be understood by including the effect of the spatial inhomogeneity.

As the temperature decreases, the total volume of FM remains unchanged while two distinguished magnetic regions appear [22,23]. In this letter, we refer to the metallic-like region as FMM region, and the insulating region as FMI region. The weak variation of the HHG dependence on MIR light intensity, with temperature around $T_{C1}$ (as plotted in Figs. 4(a) and 4(b)), indicates the insulating phase is the main contributor to the HHG signal over the whole temperature range. Accordingly, the nonlinear response of FMM is likely to be significantly weaker than that of FMI. The difference in HHG between the metallic and insulating phases was previously studied on a strongly correlated material $VO_2$, and our hypothesis is consistent with their results [12].

To take into account the effect of the spatial inhomogeneity, there are two possible processes. One is that the existence of FMM with low HHG efficiency causes a reduction in HH intensity. A study on negative differential resistance shows that at low temperatures, a smaller bias current is needed to induce carrier delocalization via DE, opening the conduction path [22]. Their results suggest that below $T_{C1}$, metallic FM correlation is enhanced as the temperature is lowered. The process where the amount of FMM increases at low temperatures and suppresses the total HH intensity seems reasonable. However, it's not sufficient for a comprehensive interpretation of our results. This is because the change ratio of HH yields below $T_{C1}$ should be proportional to volume fraction of FMI phase, but those of the third (Fig. 3(a)) and fifth harmonic yields (Fig. 3(b)) are not the same.

Another possible process is that because of the existence of randomly distributed FMM regions, the strength and phase of the electric field acting on FMI regions is spatially inhomogeneous. Thus, destructive interference can occur when the emissions from FMI regions at different positions are added together, leading to a reduction in the overall HH intensity. The difference in the behavior of third and fifth harmonics can be interpreted as their different sensitivity to spatial inhomogeneity owing to its different emission wavelength. Further wavelength-dependent measurement may clarify the effect of inhomogeneity on HH process. The spatial inhomogeneity arises from the random distribution of $Mn^{3+}$ and $Mn^{4+}$. Thus, this process is peculiar for non-charge-ordered PCMO systems.

In addition to the thermal fluctuation of orbital order, the contribution of the intrinsically inhomogeneous ferromagnetic ground state is also incorporated for a complete understanding of the experimental data. Compared to already reported results on HHG from CO/OO manganite

$Pr_{0.6}Ca_{0.4}MnO_3$ [15], the experimental features observed here are similar in OO phase and different in spin-ordered phase. We attributed the newly observed behavior in the spin-ordered phase to the spatial inhomogeneity in FMI manganites.

## V. SUMMARY

In summary, HHG measurement has been performed on a ferromagnetic insulating manganite $Pr_{0.8}Ca_{0.2}MnO_3$. Temperature-dependent HHG measurement has allowed us to obtain information on the nature of the strong correlation effect on HHG process. Above $T_{C1}$, thermal fluctuation of orbital order plays a key role. Below $T_{C1}$, the contribution of the intrinsically inhomogeneous ground state, where weakly linked metallic-like regions exist in a predominantly insulating medium, must be taken into account. This study shows that HH intensity is extremely sensitive to spatial inhomogeneity caused by strong electron correlation. This is in contrast to other spectroscopic methods such as absorption spectroscopy, and demonstrates the effectiveness of HHG spectroscopy in strongly correlated electron systems [19].

**Acknowledgments**

This work is supported by Grants-in-Aid for Scientific Research (S) (Grant No. JP21H05017, No. JP17H06124) and Grants-in-Aid for Scientific Research (C) (Grant No. JP22K03484). A. N. is thankful for JST SPRING, Grant No. JPMJSP2110. K. U. is thankful for a Grant-in-Aid for Challenging Research (Pioneering) (Grant No. 22K18322).

Figures & Captions

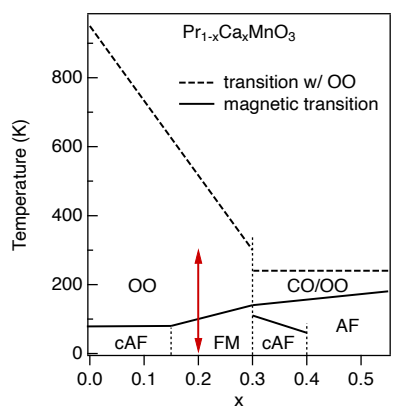

FIG. 1. Composition-temperature phase diagram of $Pr_{1-x}Ca_xMnO_3$ [16,17]. Dashed lines indicate the phase transition with orbital ordering (OO) of Mn $e_g$ electrons. For $x \geq 0.3$, charge ordering (CO) also appears. Solid lines indicate magnetic critical points (Néel temperature $T_N$, or Curie temperature $T_C$). AF, cAF, and FM denote antiferromagnetic, canted antiferromagnetic, and ferromagnetic phases, respectively.

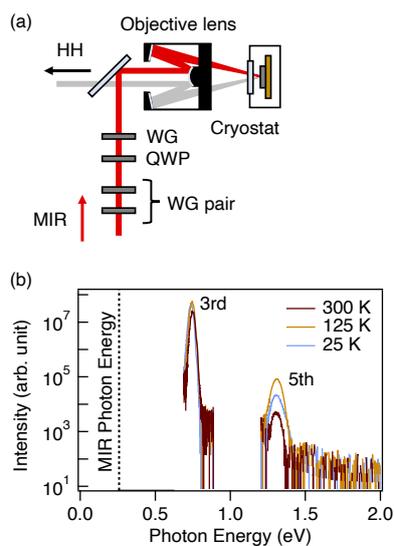

FIG. 2. (a) Schematics of experimental setup. WG and QWP denote wire-grid polarizer and quarter-wave plate. (b) High harmonic spectra at different temperatures, 25 K, 125 K, and 300 K. Two peaks are observed, corresponding to third and fifth harmonics of incident light. Dashed line shows the photon energy of the incident MIR light, 0.26 eV.

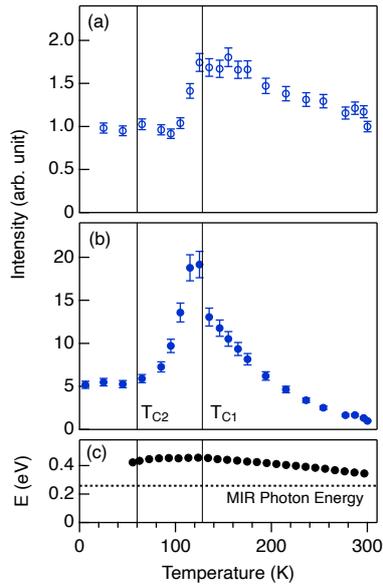

FIG. 3. Temperature variation of (a) third and (b) fifth harmonic intensity. (c) Temperature variation of optical gap energy following Ref. [19]. The gap energy is larger than the photon energy of incident MIR light (plotted by dashed line) at entire temperature range.

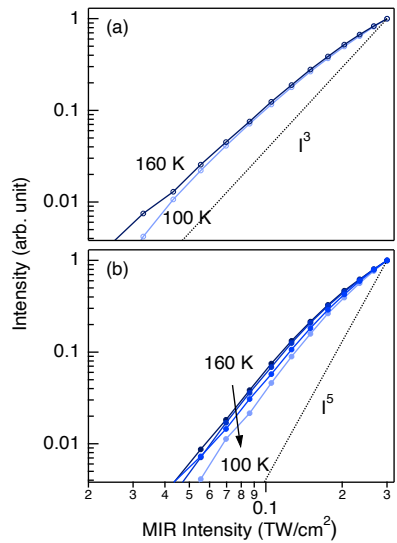

FIG. 4. MIR intensity dependence of (a) third harmonic intensity and (b) fifth harmonic intensity at several temperature points around $T_{C1} \approx 130$ K.

# Supplemental Material for "Nonlinear optical response in a ferromagnetic insulating manganite: $Pr_{0.8}Ca_{0.2}MnO_3$"

Temperature dependence of reflectivity at 0.26 eV

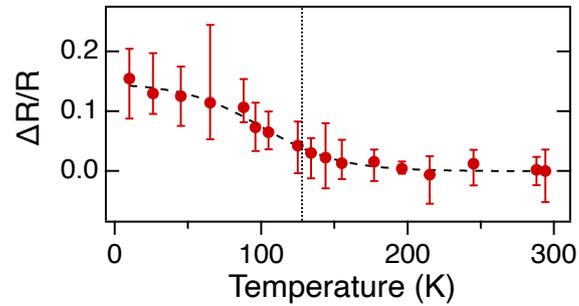

Reflectivity at the photon energy of incident mid-infrared light (0.26 eV) was measured within the temperature range of 10 K ∼ 300 K. The temperature evolution is monotonic and insufficient to explain the peak structure around $T_{C1} \approx 130$ K in the temperature variation of high-harmonic intensity as plotted in Figs. 3(a) and 3(b) in the main text.